\documentclass{aa}
\usepackage{graphicx,pifont}
\usepackage{epsfig}
\usepackage{psfig}
\usepackage{amssymb}
\usepackage{latexsym}
\usepackage{longtable}
\usepackage{subfigure}
\usepackage{amsmath}
\usepackage{array}
\usepackage{txfonts}
\begin{document}
    \title{Asiago eclipsing binaries program. II. V570 Per}

\author	{P.M. Marrese \inst{1,2}
	\and
	U. Munari\inst{2,3}
	\and
	R. Sordo\inst{2}
	\and
	S. Dallaporta\inst{4}
	\and
	A. Siviero\inst{1,2}
	\and
	T. Zwitter\inst{5}
          }
   \offprints{U.Munari, {\tt munari@pd.astro.it}}

\institute{
	Dipartimento di Astronomia dell'Universit\`a di Padova,
	Osservatorio Astrofisico, 36012 Asiago (VI), Italy
        \and
	Osservatorio Astronomico di Padova, Sede di Asiago, 36012
	Asiago (VI), Italy
	\and
	CISAS, Centro Interdipartimentale Studi ed Attivit\`a Spaziali
	dell'Universit\`a di Padova             
	\and
	Via Filzi 9, I-38034 Cembra (TN), Italy
	\and
	University of Ljubljana, Department of Physics, Jadranska 19, 1000
	Ljubljana, Slovenia
        }

   \date{Received date ................; accepted date ...............}

   \abstract{The orbit and physical parameters of the double-lined eclipsing
     binary V570~Per, discovered by the Hipparcos satellite, are derived
     with formal errors better than 1\% using high resolution Echelle
     spectroscopy and $B$, $V$ photometry.  Atmospheric analysis is
     performed on spectra at quadrature using synthetic spectroscopy that indicates a
     [Fe/H]=+0.02$\pm$0.05 metallicity. V570 Per turns out to be a detached
     system, with shallow eclipses of $\Delta B$=$\Delta V$=0.13~mag and
     a distance of 123$\pm$3~pc, in fine agreement with Hipparcos
     117$^{104}_{132}$ pc 
     distance. V570~Per is composed by unperturbed components of F2 and F5
     spectral types and masses of 1.457$\pm$0.004 and 1.351$\pm$0.004
     M$_\odot$ respectively, which do not show surface activity.  Both
     components are still within the Main Sequence band and are dynamically
     relaxed to co-rotation with the orbital motion. The system is
     particularly interesting because both components have their masses in
     the range where transition occurs between convective and radiative
     cores, and where differences between families of stellar evolutionary
     tracks show appreciable differences. The position of the
     components of V570~Per on the temperature-luminosity plane is compared
     with Padova, Geneva, Granada and Teramo-04 theoretical stellar models.
     The comparison provides different best fit metallicities and ages, with
     [Fe/H] values ranging from +0.07 to +0.18, and ages from 0.6 to
     1.0 Gyr.

            }

   \maketitle

\section{Introduction}

The aim of this series of papers is to contribute to the determination of high
precision masses, radii and temperatures of stars with high quality orbital
solutions for eclipsing binaries and to compare with the predictions of theoretical
stellar models. Our targets have spectral types F, G or K as their paucity
among the well determined eclipsing systems was underlined by Andersen (1991,
2002).  Siviero et al. (2004), hereafter Paper I, outlines in details the type
of data and methods used throughout this series of papers.

V570~Per (HD 19457, HIP 14673) is a nearby eclipsing binary of early F spectral
type, with a period of 1$\overset{\rm d}{.}$9 which was discovered by the
Hipparcos satellite. As it will be seen in the next sections the system has a
well detached nature and no intrinsic variability, which make it a useful
object to be compared with the theoretical evolutionary models of single stars.
V570~Per is seen projected toward the $\alpha$~Per (Melotte~20) young open
cluster ($\alpha$=03~22.0, $\delta$=+48~37), but it is a foreground star and
thus is not physically associated with the cluster.  This can be deduced from
the comparison between the distance and proper motion of $\alpha$~Per
(d=183$^{177}_{190}$ pc, $\mu_\alpha^*$=$+$22.47$\pm$0.16~mas yr$^{-1}$,
$\mu_\delta$=$-$25.99$\pm$0.17~mas yr$^{-1}$, from Van Leeuven 1999) and of
V570~Per (d=117$^{104}_{132}$ pc, $\mu_\alpha^*$=$+$52.20$\pm$0.85~mas
yr$^{-1}$, $mu_\delta$=$-$41.58$\pm$0.797~mas yr$^{-1}$, from Hipparcos
Catalogue). In this paper a simultaneous photometric and spectroscopic orbital
solution of V570~Per is presented. It is based on high resolution ($R_{\rm
P}$=20\,000), high S/N ($\geq$~100) Echelle spectra covering the wavelength
region $\lambda\lambda$ 4500$-$9480~\AA\ and high precision photometry in the
Johnson $B$ and $V$ bands. We obtain accurate orbital and stellar parameters,
perform an atmospheric analysis via comparison with synthetic Kurucz spectra
and determine the evolutionary status and age through comparison with
theoretical stellar models and isochrones. We use the comparison between the
distance determined by our orbital solution and the Hipparcos parallax to
validate the orbital analysis.

A preliminary photometric and spectroscopic study of V570~Per is present in
literature (Munari et al. 2001) as part of the simulation of the performances
on eclipsing binaries of the ESA's mission GAIA. This previous solution relies
on Hipparcos/Tycho photometry and on ground based radial velocities (obtained
in the $\lambda\lambda$~8490$-$8750 \AA\ GAIA wavelength range) of lower
accuracy than the ones used here. The broad system properties derived by Munari
et al. (2001) are in agreement with those derived in this paper.

\section{The data}

\subsection{Photometry}

    \begin{figure*}
    \centerline{\psfig{file=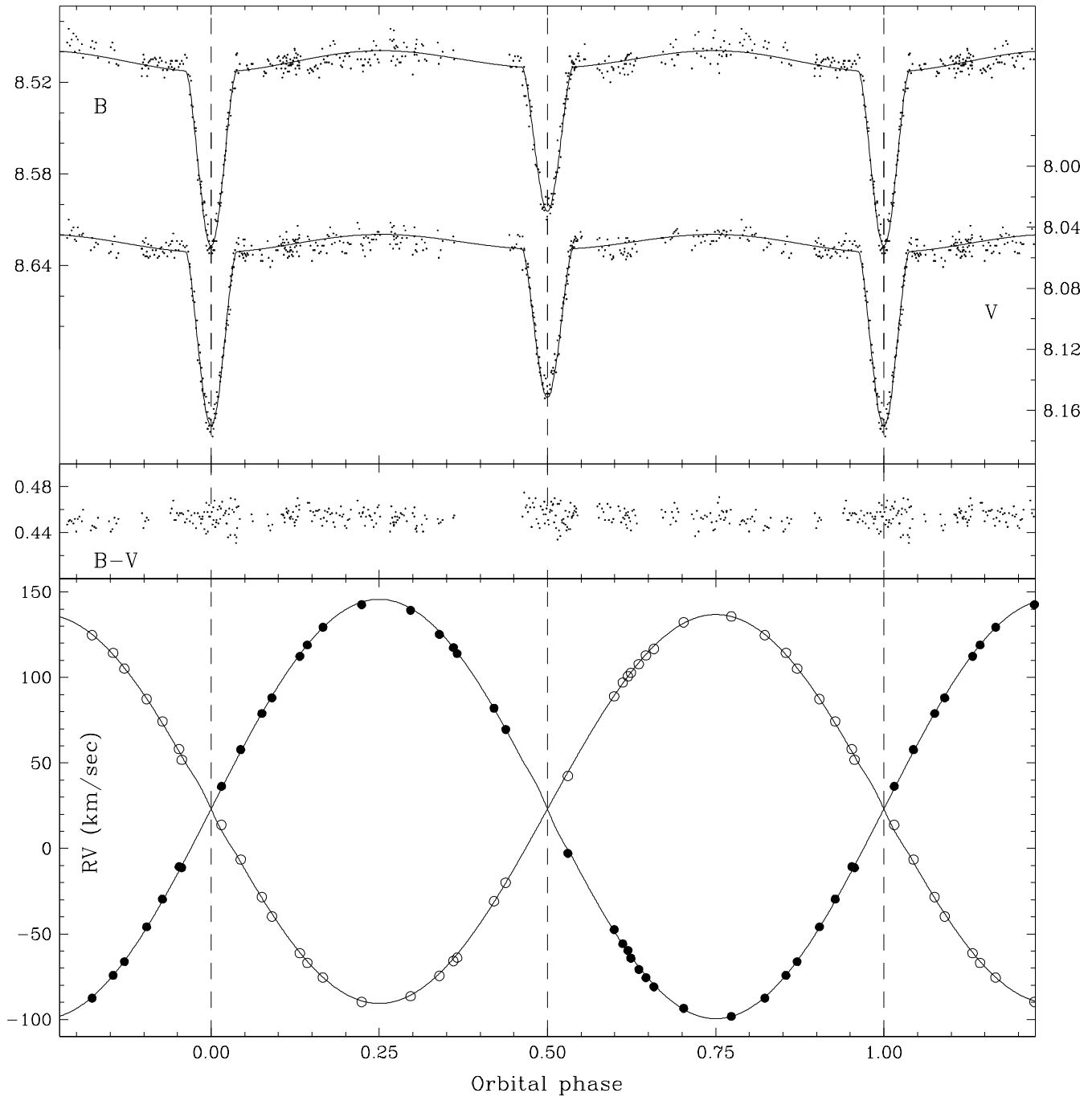,width=18.0cm}}
    \caption{The observed $B$, $V$, {\em B$-$V} and radial velocity curves of
	V570~Per. In the radial velocity panel, the open circles indicate the 
	hotter and more massive (primary) star, while the filled circles pertain to
	the cooler and less massive (secondary) star. The orbital solution
	is over-plotted to the observed data.}
    \end{figure*}

    \begin{table}
    \caption{Radial velocities of V570~Per. The columns give the spectrum number
    (from the Asiago Echelle log book),
    the heliocentric JD ($-$2451000), the orbital phase, the radial velocities of the two
    components and the corresponding errors, and the $<$S/N$>$ of the spectrum
    averaged over the wavelength range considered in the analysis.}
    \begin{center}
    \centerline{\psfig{file=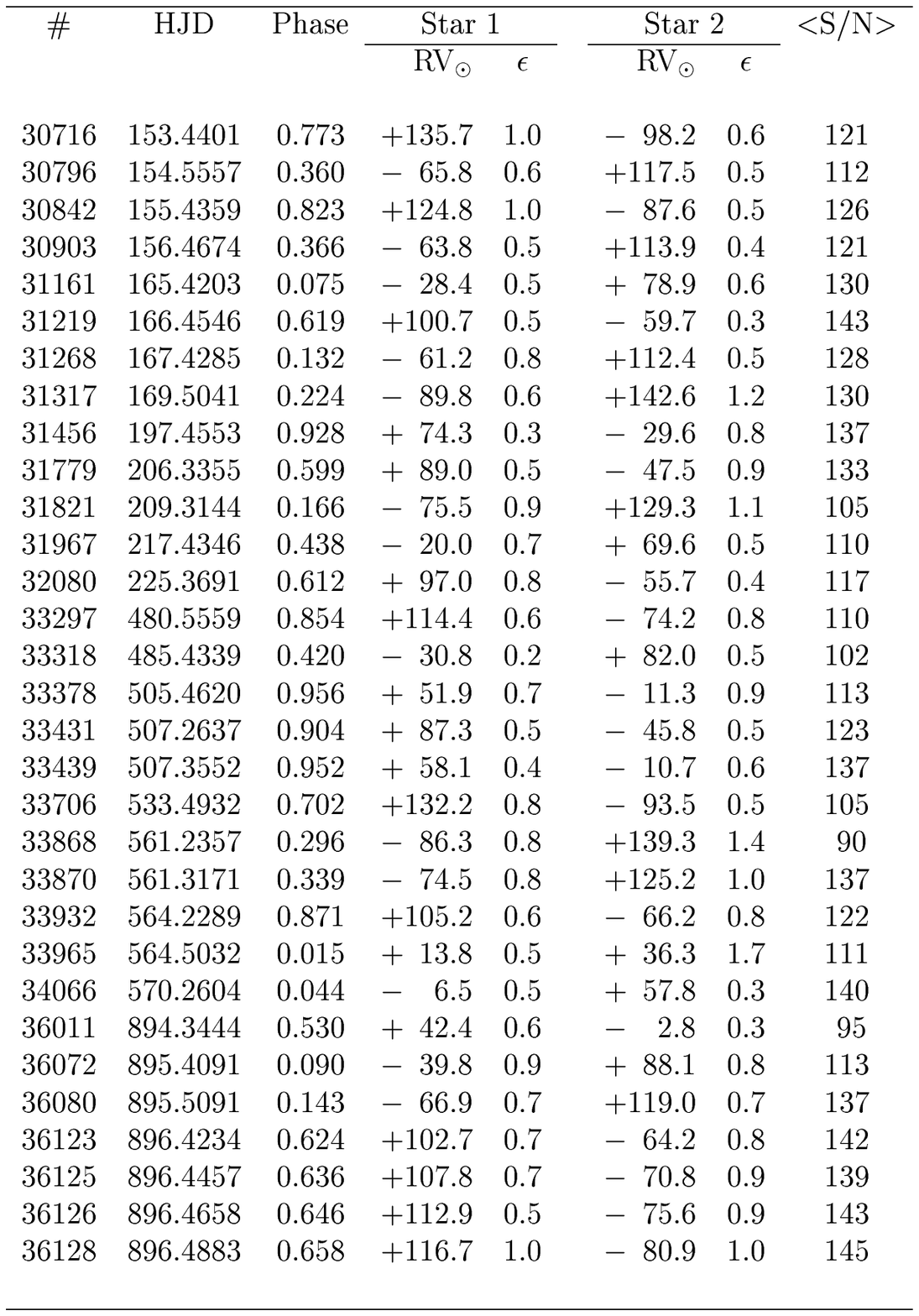,width=8.7cm}}
    \end{center}
    \end{table}

The photometric observations of V570~Per were obtained in $B$ and $V$ (standard
Johnson filters) from a private observatory near Cembra (Trento),
Italy. The instrument is a 28 cm Schmidt-Cassegrain telescope
equipped with an Optec SSP5 photometer. The diaphragm has a size of
77 arcsec. There are no
stars in the aperture brighter than $V$=15 mag that could even minimally
interfere (i.e. all possible field stars in
the aperture are al least 7 mag fainter than V570~Per).
The instrumentation already proved to be very
accurate and reliable (cf. Paper~I) and thus perfectly suited to
deal with the low amplitude eclipses of V570~Per ($\sim$0.13 mag in
both $B$ and $V$ bands).
 
The comparison star is HD 19805 (HIP 14980, 
$B_{\rm T}$=8.108$\pm$0.015, $V_{\rm T}$=7.973$\pm$0.012, 
spectral type B9.5~V) and the check star is TYC~3315~308~1 
($B_{\rm T}$=9.919$\pm$0.029, $V_{\rm T}$=9.567$\pm$0.032). 
Both the comparison and check
stars are close to V570~Per on the sky (the distances being $\sim$42
and $\sim$18 arcmin respectively) so the atmospheric corrections are
rather small. All the observations were obtained for zenith distances
smaller than 60$\degr$, thus providing a high internal consistency to our 
photometry of V570~Per.
The comparison star was measured against the check star at least once
every observing run. In all, 34 measurements of the magnitude difference
comparison$-$check star were collected, providing a constant magnitude
difference with a standard deviation of 0.006 mag. Our results
confirm the Hipparcos/Tycho findings that both the
comparison and the check stars are not variable, and thus well
suited to serve in the photometry of V570~Per. Following the
Bessell (2000) transformations between Tycho and Johnson photometric
systems, we adopted $B$=8.073 and $V$=7.957 for the comparison star.

In all, 446 measurements in $B$ and 465 in $V$ were collected of V570~Per
between 2000 and 2003\footnote{Data at {\tt
http://ulisse.pd.astro.it/Binaries/V570\_Per/}}. Each photometric point is
actually the mean of 10 consecutive and independent measurements (each one 5
sec long) and the typical error of the mean for each photometric point is
0.006 mag in $B$ and 0.005 mag in $V$. All the observations are corrected
for atmospheric extinction and color corrections (via calibration on
Landolt's equatorial fields), and the instrumental differential magnitudes
are transformed into the standard Johnson UBV system.

The light curves of V570~Per in each band as well as the ({\em B$-$V}) color are 
shown in Figure~1. The observations have a reasonably good phase coverage 
(except the phase immediately preceding the secondary eclipse). As 
mentioned above the eclipses are very shallow due to the low inclination of 
the orbit ($i$=77$\degr$ in our solution). This is the main reason for 
the absence of color variations during the eclipses (both stars remain essentially 
visible throughout the whole eclipses). The mean values out of eclipses
are $B$=8.505 and $V$=8.052 mag. 
The dispersion of the $B$ and $V$ measurements 
about their mean value out of the eclipses ($\sigma_B$=0.007, 
$\sigma_V$=0.006) is only marginally higher than the accuracy of a single 
measurement. Thus any intrinsic variability of an amplitude larger than 0.007 mag 
should be ruled out.

    \begin{figure}
    \centerline{\psfig{file=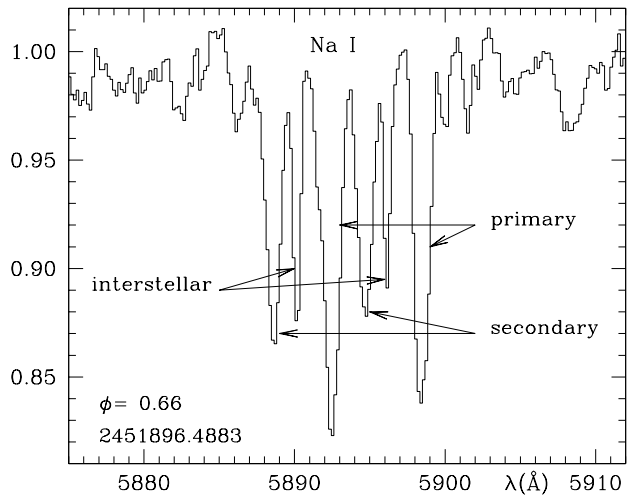,width=8.0cm}}
    \caption{The NaI doublet (5890 \& 5896 \AA) region for V570~Per. The 
	interstellar components are clearly separated from the stellar ones.
	The equivalent width of the interstellar NaI ($\lambda$~5890) is 0.08$\pm$0.03~\AA\
	which correspond to $E_{B-V}\sim$0.02 as described in Section 4.}
    \end{figure}

\subsection{Spectroscopy}

The spectra of V570~Per were obtained in 1999$-$2002 with the Echelle+CCD 
spectrograph on the 1.82 m telescope operated by Osservatorio Astronomico 
di Padova atop Mt.Ekar (Asiago). The instrumentation and observing set-up exactly match
those described in Paper I, to which we refer for details of the observing mode. Here we recall 
that the wavelength region covered is $\lambda\lambda$~4500$-$9480~\AA\, 
with a resolving power $R_{\rm P}$~$\sim$~20\,000.
A journal of the observations is given in Table~1.
The observations were planned so as to obtain a good phase coverage (cf. Figure~1).
In all we obtain 31 spectra with exposure times from 1200 to 1800 seconds, which 
guarantee a good S/N ratio (cf. Table~1) while preserving from the smearing due to
the orbital motion (1500 seconds correspond to less than 1\% of the orbital period). 

\section{Radial velocities}

In deriving the radial velocities of V570~Per we follow strictly the same
strategy outlined in Paper I, where an accurate and detailed description of
the method is given. In short, six adjacent Echelle orders covering entirely
the wavelength range from $\lambda$~4890 to $\lambda$~5690~\AA\ are
measured via a two-dimensional cross-correlation technique (TODCOR) based on
the Zucker \& Mazeh (1994) algorithm. These six Echelle orders are chosen
because they are densely populated by absorption lines, in particular FeI
and MgI, that perform particularly well in terms of radial velocity. As
templates we use synthetic spectra with the appropriate temperatures,
surface gravities and rotational velocities. The templates are selected
from the large synthetic spectral atlas computed at the same 20\,000
resolving power with Kurucz's codes by Munari et al. (2004, in submission).

The results of the radial velocity measurements are summarized in Table~1.
The mean error of radial velocities is 0.6~km~sec$^{-1}$ for star~1, and
0.7~km~sec$^{-1}$ for star~2, as estimated from comparison of the radial
velocities obtained separately from each of the six Echelle orders analyzed
here.

\section{Reddening determination}

The measurement of reddening is a key step in the determination of the
absolute temperature scale (and therefore of the distance) of eclipsing
binaries. In spite of its short distance, some reddening is expected to
affect V570~Per given its low galactic latitude ($l$=145.18, $b$=$-$8.19).

Our spectra cover the interstellar NaI (5890 \& 5896 \AA) and KI (7665 \&
7699 \AA) doublets which are excellent estimators of the reddening as
demonstrated by Munari \& Zwitter (1997). They calibrated a tight relation
between the NaI D2 ($\lambda$5890~\AA) and KI ($\lambda$7699~\AA) equivalent
widths and the color excess $E_{B-V}$. In 16 out of 31 of our spectra
the orbital motion separates unambiguously the stellar lines from the
interstellar components ($\Delta(\lambda_{\rm star} - \lambda_{\rm
ISM})\geq$ 2 FWHM) allowing a safe measurements of the equivalent widths of the interstellar lines
(see Figure~2 where an example of the NaI doublet region for V570~Per is
shown). We obtain for the equivalent width of NaI~($\lambda$~5890)=0.08$\pm$0.03~\AA.
This corresponds to $E_{B-V}\sim$0.02. The absence of any detectable
KI line in our high resolution high S/N spectra confirms the very low
reddening affecting V570~Per.

   \begin{table}
    \caption{Orbital solution for V570~Per (over-plotted to observed data in
    Figure~1). Formal errors to the solution are given. The last two lines
    compare the Hipparcos trigonometric parallax (and its 1$\sigma$ error
    interval) with the distance derived from the orbital solution.}
    \begin{center}
    \centerline{\psfig{file=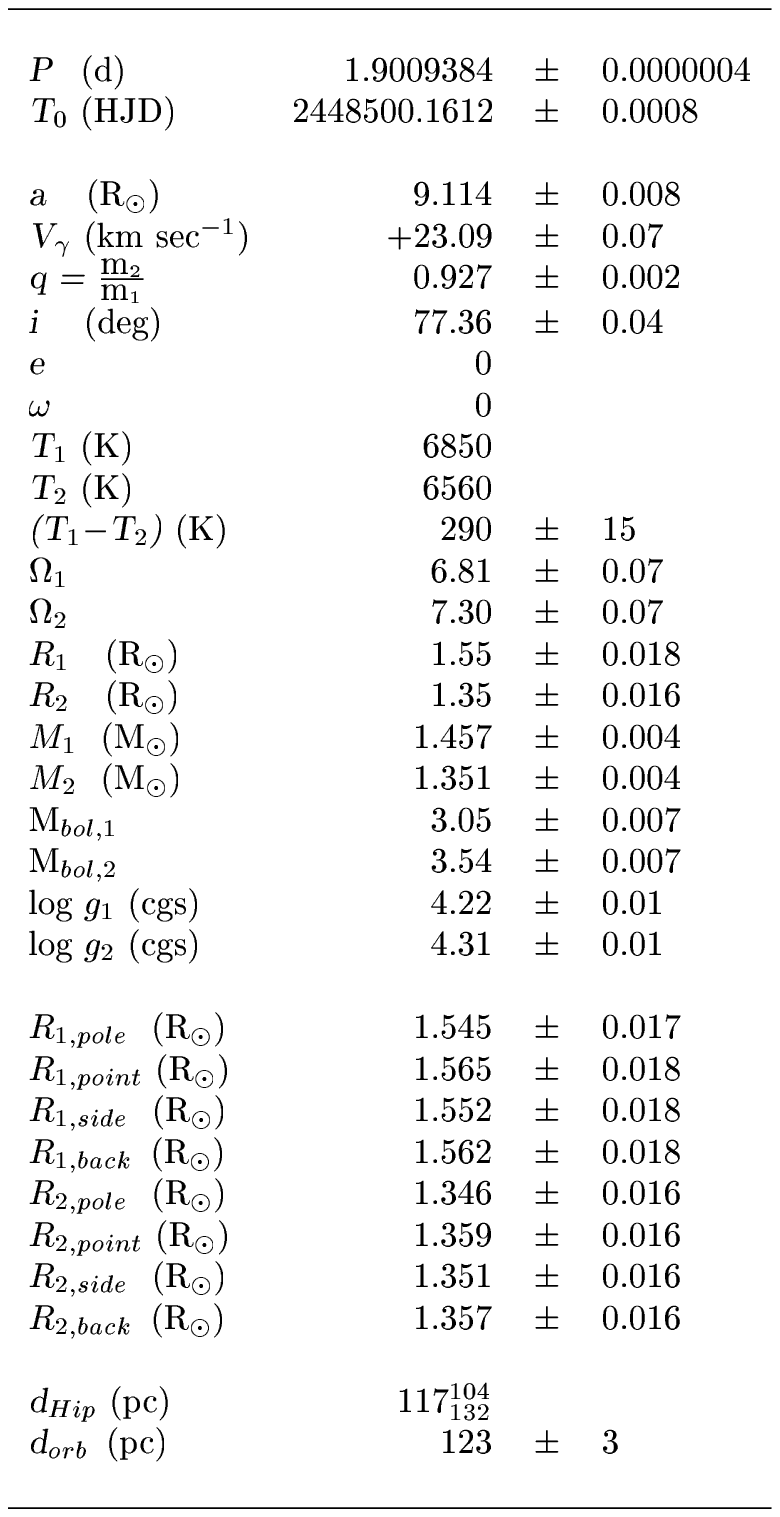,width=6.7cm}}
    \end{center}
    \end{table}

Estimating the reddening from the equivalent widths of NaI and KI interstellar lines is a
well established method. There are some sources in literature which
confirm the very low reddening we found for V570~Per.
As mentioned in the introduction, V570~Per is seen in the direction of the
$\alpha$~Per cluster, even if it is a foreground star, and the reddening
affecting $\alpha$~Per is thus an upper limit for V570~Per. The $\alpha$~Per
cluster extends over a wide area on the sky and presents a patchy
distribution of the reddening over it. According to Crawford \& Barnes
(1974) its mean reddening is $E_{B-V}$=0.09, while Prosser (1992) find
$E_{B-V}$=0.11. It is also worth mentioning that HD~19665, which lies less
than 1 arcmin from V570~Per, has been measured by Perry \& Johnston (1982),
who found $E_{B-V}$=0.022. Noting that HD~19665 is somewhat farther away than
V570~Per ($d$=150$^{130}_{178}$~pc, from Hipparcos) this is an excellent
confirmation of the $E_{B-V}$=0.02 we found for V570~Per.

\section{Orbital solution}
\subsection{Initial guess of the orbital and stellar parameters and modeling strategy}

The orbital modeling is performed with the Wilson-Devinney code 
(Wilson \& Devinney 1971, Wilson 1998) with modified stellar 
atmospheres $WD98K93$ (Milone et al. 1992) and with the limb
darkening coefficients from van Hamme (1993). 

First of all we obtain an initial guess of the temperature of the primary
($T_{\rm 1}$), in other words of the reference temperature.  We do this
independently from the results of the atmospheric analysis via synthetic
Kurucz spectra, to the aim of using the latter analysis as an external
check. From our photometry and our reddening determination we obtain for
V570~Per out of eclipses ({\em B$-$V})$_{\circ}$=+0.45$-$0.02~=+0.43. An
inspection of the spectra and of the eclipse depths tells us the two stars
are similar. Supposing both stars have equal temperatures, the de-reddened
color suggests (according to Fitzgerald 1970 and Popper 1980 conversion
tables) that they have an F4-F5 spectral type. We confirm this result by
comparing the spectrum of V570~Per in the
$\lambda\lambda$~~8480$-$8740~\AA\, GAIA wavelength region with the spectral
atlases Munari \& Tomasella (1999) and Marrese et al. (2003), which cover
the same interval at the same resolution.  Using the calibrations from
Strai\v zys \& Kuriliene (1981) an F4-F5 spectral type corresponds to an
effective temperature of 6700~K.  We thus initially fix $T_{\rm 1}$=6700~K
and from the difference of eclipse depths, which gives the ratio of stellar
temperatures, we determin the initial value of $T_{\rm 2}$. The GAIA-like
orbital solution of Munari et al. (2001) is used as the set of initial
guesses for the other parameters (period $P$, epoch $T_{\rm 0}$, semi-major
axis $a$, barycentric radial velocity $V_{\rm \gamma}$, mass ratio
$q=M_1/M_2$, inclination $i$, eccentricity $e$, modified Kopal potentials
$\Omega_{\rm 1,2}$ and relative luminosity in each passband of the primary
star L$_1$). In addition to our 2000$-$2003 photoelectric photometry, the
Hipparcos data in $B_{\rm T}$, $V_{\rm T}$ and $H_{\rm P}$ bands (covering
the time interval 1990$-$1993) are used to adjust the period $P$ and the
epoch $T_{\rm 0}$ so as to increase the time span. We find no evidence for
a change in the orbital period and a formal upper limit for dP/dt of 0.0025
sec~yr$^{-1}$.

We compute the solution using mode 2 of the Wilson-Devinney code, which is
appropriate for detached binaries with no constraints on the potentials. As
the potentials of both stars are permitted to vary, the stellar sizes are
constrained by the eclipses alone. The relative luminosity of the secondary
star is calculated from the other parameters ($L_{2}$ being coupled with the
temperatures). We use a square root limb darkening law with the coefficients
interpolated from van Hamme (1993) tables for the appropriate
temperatures and surface gravities. In doing this we assume a solar metallicity
and this is later and independently confirmed by the atmospheric analysis.
The bolometric albedo $A$
and gravity brightening coefficients $\beta$ are fixed to unity values
appropriate for radiative atmospheres. V570~Per light and radial velocity
curves do not show a clear signature of an eccentric orbit. The circularity
of the orbit is confirmed by initial modeling runs during which
eccentricity is allowed to vary and nevertheless remains consistent with
zero. After a few of such trials $e$ is set to zero.

In the explicit standard use of the Wilson-Devinney code, the light curves are treated as flux values normalized
to unity out of eclipses, separately for each photometric band. This causes a loss of 
information on the color temperature of the stars and therefore the necessity to assume a
temperature for the primary ($T_{\rm 1}$). However, the limb darkening coefficients do depend on the 
absolute temperature. The goodness of the choice for $T_{\rm 1}$ can be assessed by
deriving a series of solutions where different values for $T_{\rm 1}$ are adopted
and by selecting the one giving the smallest residuals. This method confirms our choice for 
$T_{\rm 1}$ within its sensitivity of $\sim$100~K.

\subsection{Orbital solution and physical parameters of the components}
 
The results of the simultaneous photometric and spectroscopic solution and 
corresponding physical quantities are given in Table~2, while the model 
radial velocity and light curves are over-plotted to the observational data in Figure~1.
The formal accuracy of the solution is 0.3\% on the masses and $\sim$1\% on the radii for both 
components. The system parameters which are mainly dependent on the radial velocity 
curves (i.e. $a$, $M_{\rm 1}$, $M_{\rm 2}$, $q$, $V_{\rm \gamma}$) are very well 
constrained. The radii, which are dependent on the eclipse fitting,
even if determined with a formal high accuracy, anyway suffer from the low inclination 
of the orbit and the consequent shallowness of the eclipses.
No departure from spherical symmetry is found ($R_{pole}$/$R_{point}$~$\sim$~0.99 for 
both stars) and the system is detached ($R_{\rm 1}$/$a$~$\sim$~0.17).

The orbital solution gives a difference in the temperatures of the two components of
290~K with a formal uncertainty of 15~K. However neither $T_{\rm 1}$ nor $T_{\rm 2}$
are known with this accuracy. We remember that $T_{\rm 1}$ is assumed to be 
6850$\pm$100~K.

To check the accuracy of the derived orbital solution, we compare
the corresponding distance with the Hipparcos parallax. In calculating the distance, 
we assume the usual relation $R$=$A_V$/$E_{B-V}$=3.1.
Our determination of the distance, d=123$\pm$3 pc, is in very good agreement with the 
Hipparcos parallax (d=117$^{104}_{132}$ pc, 1$\sigma$ limits).

    \begin{figure*}
    \centerline{\psfig{file=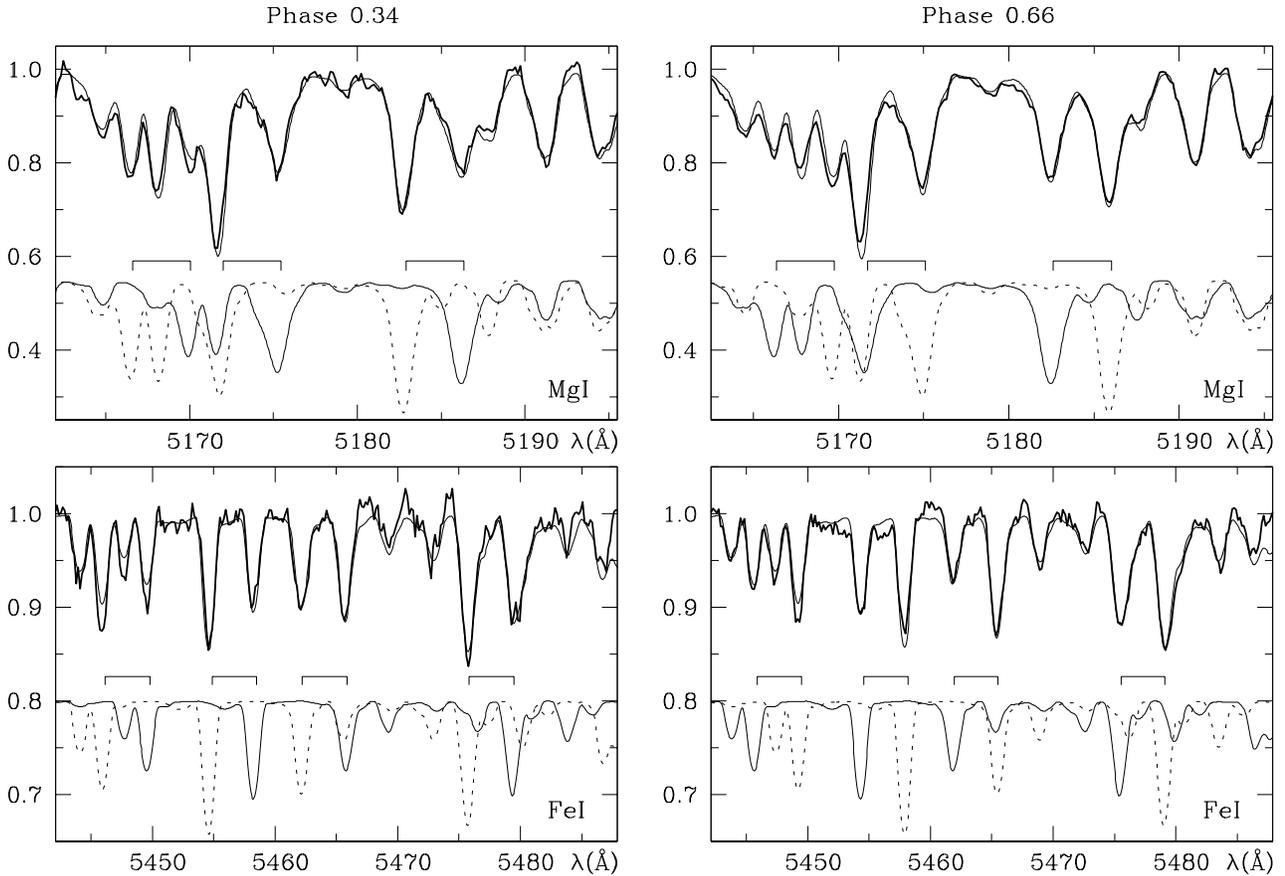,width=17.6cm}}
    \caption{Comparison between observed (thick line) and synthetic (thin line)
    V570~Per spectra over two sample wavelength regions dominated by MgI lines
    (top) and FeI lines (bottom). Spectra at orbital phases 0.34 and 0.66 (\#
    33870 and 36128 in Table~1, respectively) show full split between the two
    components and their position interchange allow a careful check of the fit
    accuracy. In each panel the lower curves represent (not to scale with the
    main spectra but in correct proportion between them) the contribution of
    each component of the binary to the formation of the observed spectrum at
    the given phase. The markers connect the shifted wavelengths of the same
    MgI (top panels) and FeI (bottom panels) lines in the spectra of the
    two components of the binary.}
    \end{figure*}

\begin{table}
    \caption{Atmospheric parameters of V570~Per from a $\chi^2$ fit to
    Kurucz synthetic spectra. The results from the orbital solution
    in Table~2 for $T_{\rm eff}$, $\log g$ and co-rotation velocity 
    are given for comparison.}
    \begin{center}
    \centerline{\psfig{file=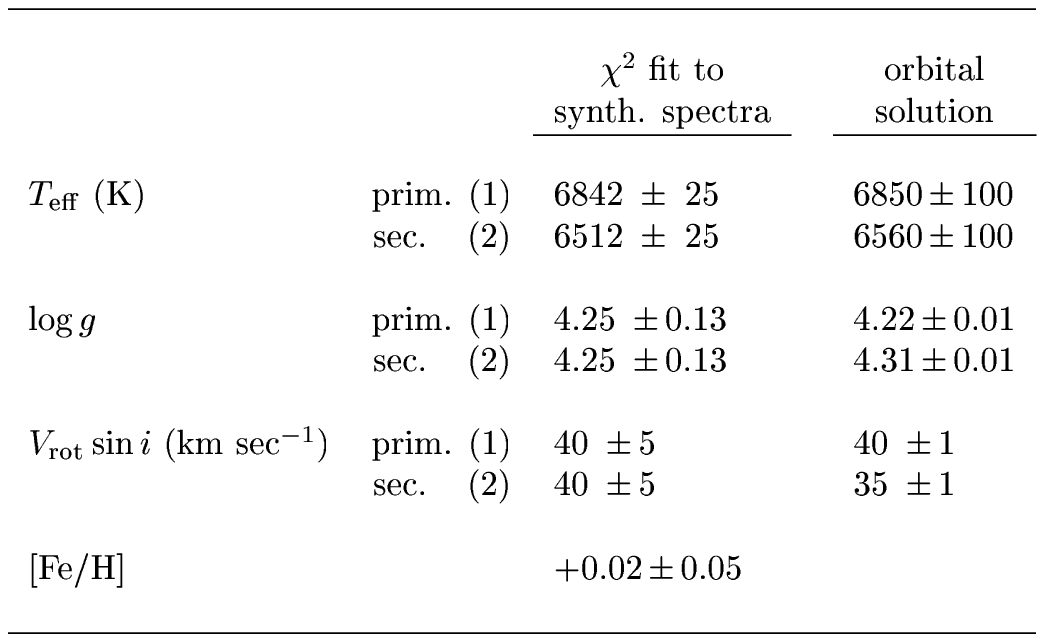,width=8.5cm}}
    \end{center}
    \end{table}

\section{Analysis of stellar atmospheres}

The spectral analysis of V570~Per allows to obtain an independent
determination of the atmospheric parameters of both components: $T_{\rm eff}$, 
$\log g$, $V_{\rm rot}$~$\sin i$ and, most important, the metallicity [Fe/H] and
microturbulent velocity, which cannot be determined or guessed from the orbital analysis.

The analysis is performed via $\chi^2$ fitting procedure, using the large grid
of synthetic Kurucz spectra of Munari et al. (2004, in submission). The
spectra cover the 2500--10\,500 \AA\ wavelength range with the same resolving power
of the observed spectra ($R_{\rm P}~$=~20\,000). The spectra are calculated with the 
revised solar abundances by Grevesse \& Sauval (1998) and the new opacity
distribution functions (ODFs) of Castelli \& Kurucz (2004),
throughout the whole HR diagram for 12 different rotational velocities, $T_{\rm eff}$
ranging from 3500 to 47\,500 K, $\log g$ from 0.0 to 5.0 and [Fe/H] from $-$2.5
to +0.5. The grid includes also a complete set of spectra calculated for
$\alpha$-enhanced chemical composition ([$\alpha$/Fe]=+0.4) and for
different values of micro-turbulent velocity (1, 2, 4 km~sec$^{-1}$).
As discussed by Zwitter et al. (2004), the grid steps are small enough to allow 
a safe interpolation between adjacent spectra.

The $\chi^2$ fitting procedure is performed on two high S/N spectra of
V570~Per taken at different phases, near 0.34 and 0.66, when the difference
in radial velocity between the components is near maximum, so that the 
lines of the two components are well separated. Using spectra obtained at two
different phases allows to account for the blending of different lines from
the two components. We concentrate our analysis on the same wavelength range
4890--5590 \AA\ used to derive the radial velocities. To remain independent
from the orbital solution, besides $T_{\rm eff}$, $\log g$, $V_{\rm
rot}$~$\sin i$, [Fe/H] and micro-turbulent velocity, also the ratio of the
luminosity of the two components is considered a free parameter. 
The $\chi^2$ procedure allows a firm identification of the location, in the grid
space, of the absolute minimum. Around it the grid is interpolated to a finer
step and the position of the minimum is re-fitted and calculated to a greater 
accuracy.
The absence of a total eclipse, however, prevents us
from observing the spectrum of the occulting star alone and makes it
difficult to measure with confidence the micro-turbulent velocity, which we
find constrained between 1.7 and 2 km~s$^{-1}$. We obtain for the
metallicity [Fe/H]=+0.02$\pm$0.05. A comparison between the results obtained
with the orbital solution ($T_{\rm eff}$, $\log g$) and the atmospheric
analysis ($T_{\rm eff}$, $\log g$, $V_{\rm rot}$~$\sin i$ and [Fe/H]) is
presented in Table~3, while in Figure~3 an example of the goodness of fit is
shown.

    \begin{figure*}
    \centerline{\psfig{file=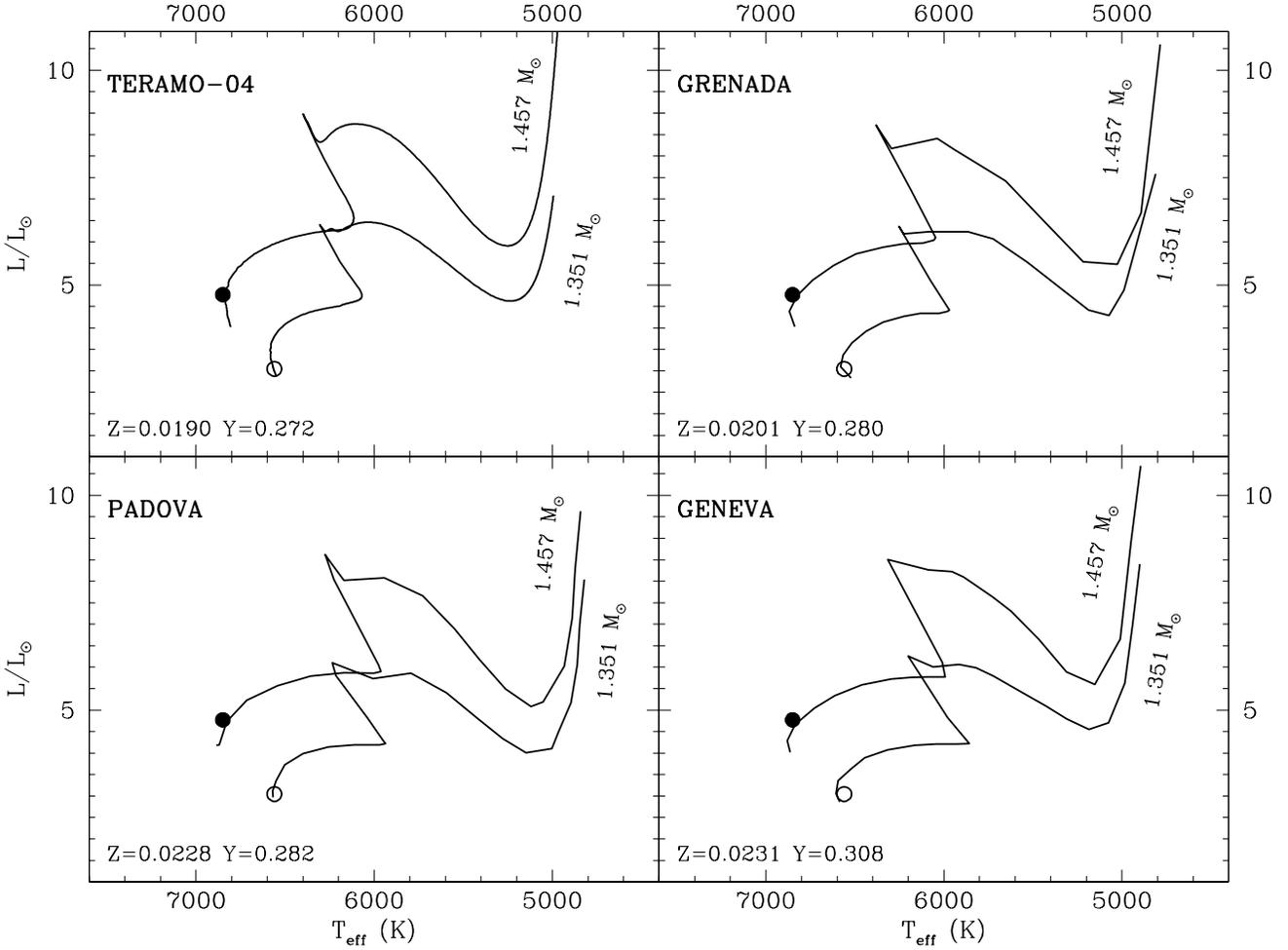,width=13.6cm,angle=270}}
    \caption{Comparison between temperature and luminosity as derived from the
    orbital solution (cf. Table~2) and those of different families of theoretical stellar 
    models for the masses of the two components of V570~Per (1.457 and 1.351 M$_\odot$)
    derived from the orbital solution.
    In each case it is possible to find two tracks
    (one for each component of the binary) which match very 
    well the observed points. However, as indicated in each panel, the
    four metal content values are markedly different and none of them matches precisely
    the metallicity derived by spectroscopic means. }
    \end{figure*}

    \begin{table}
    \caption{Metallicity for V570~Per as obtained from comparison with 
             different families of theoretical evolutionary tracks. }
    \centerline{\psfig{file=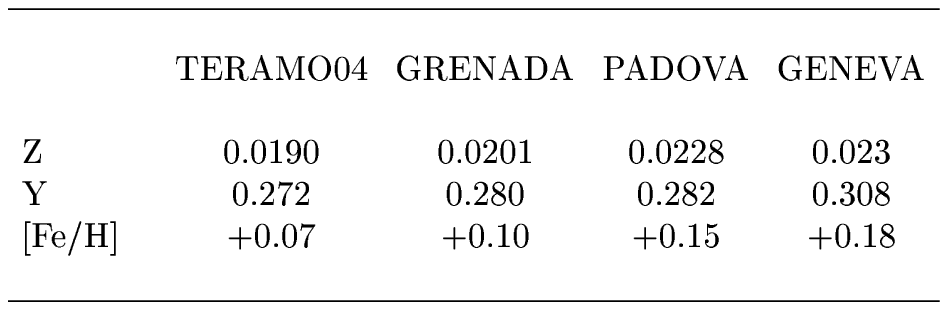,width=8.5cm}}
    \end{table}

\section{Comparison with stellar theoretical models}

The above orbital and atmospheric analyses have provided us with accurate masses, radii,
temperatures, and, assuming $T_{\odot}$=5770~K, luminosities of the components
of V570~Per, which allow us to directly place them on the theoretical $T_{\rm eff}$,
$L/L_{\odot}$ plane 

We compare on the Temperature/Luminosity plane the position of
the two components of V570~Per (from the orbital solution in Table~2) with
the tracks from the Padova (Bertelli et al. 1994, Fagotto et al. 1994, Girardi
et al. 2000), Geneva (Schaller et al. 1992, Schaerer et al. 1993a, 1993b),
Grenada (Claret 1995, 1997 and Claret \& Gimenez 1995) and Teramo-04
(Pietrinferni et al. 2004) families of stellar models. 
The models are interpolated in both mass and metal content Z by Newton 
polinomials of second order. It is worth noticing that the first Gyr 
of stellar age is coarsely mapped by the tracks (3-4 points only) and that this
prevents us from deriving the age of the system with the accuracy which our 
observational results would permit. In comparing with the theoretical
predictions we take 
the masses of the two components from our orbital solution, while we minimize
in luminosity ($L/L_{\odot}$), temperature ($T_{\rm eff}$) and age difference
between the components ($\tau_{\rm a} - \tau_{\rm b}$), by defining the
following weighting scheme for the $\chi^2$:\\
\begin{equation*}
\begin{split}
\chi^2(Z_{\rm sys},\tau_{\rm a},\tau_{\rm b})=
\frac{(\log{\tau_{\rm a}} - \log{\tau_{\rm b}})^2}{\sigma^{2}_{\log{\tau_{\rm a}}} + \sigma^{2}_{\log{\tau_{\rm b}}} }+\\
\frac{(\log{T^{\rm obs}_{\rm eff,a}} - \log{T^{\rm th}_{\rm eff,a}})^2}{\sigma^{2}_{\log{T_{\rm a}}}} +
\frac{(\log{T^{\rm obs}_{\rm eff,b}} - \log{T^{\rm th}_{\rm eff,b}})^2}{\sigma^{2}_{\log{T_{\rm b}}}} +\\
\frac{(\log{L^{\rm obs}_{\rm eff,a}} - \log{L^{\rm th}_{\rm eff,a}})^2}{\sigma^{2}_{\log{L_{\rm a}}}} +
\frac{(\log{L^{\rm obs}_{\rm eff,b}} - \log{L^{\rm th}_{\rm eff,b}})^2}{\sigma^{2}_{\log{L_{\rm b}}}} \\
\end{split}
\end{equation*}
For the families of tracks for which the surface gravity is available (Padova and Grenada),
we rerun the $\chi^2$ minimization substituting the luminosity ($L/L_{\odot}$) with the 
surface gravity ($\log{g}$), because of the correlation between $L/L_{\odot}$ and $T_{\rm eff}$.
The results we obtain are fully consistent with those obtained in the Luminonsity/Temperature plane.

The sets of stellar models considered, have different helium content Y for the 
same metal content Z. In order to compare the results from the various families
on metallicity with the metallicity obtained by the atmospherical analisys we
transform the metal content into [Fe/H] using
[Fe/H]=$\log{(Z/X)} - \log{(Z/X)_{\odot}}$, where we assume $(Z/X)_{\odot}$=0.023 from
Grevesse \& Sauval (1998).

Each panel of Figure~4 shows the best fit in metal content (Z) provided by the given 
family of stellar models after interpolation of closest published tracks. The results are 
summarized in Table~4. In each case it is possible to find two tracks
(one for each component of the binary) which match very 
well the observed points. The derived metallicities, however, show a large spread and 
cover a range from [Fe/H]=+0.07 to +0.18. In addition none of those [Fe/H] values is in
agreement with the metallicity derived by the atmospheric analysis ([Fe/H]=+0.02), 
even if they are close. The ages derived from the different families of stellar models are
in broad agreement (covering a range beteween 0.6 and 1.0 Gyr) and place both stars still 
within the Main Sequence band.

The masses of the two components of V570~Per, 1.457 and 1.351~M$_\odot$,
place them in the transition zone on the Main Sequence from fully convective
to fully radiative cores. This is a region where appreciable differences
occurs among the various families of theoretical stellar models, mainly
originating from the different ways in which the overshooting is treated and
imposed to vanish in fully radiative cores (cf. Pietrinferni et al. 2004,
Sandquist 2004). The accuracy obtained from 
the orbital solution for V570~Per on masses and radii (thanks to the high 
accuracy of both photometry and spectroscopy) and the position of V570~Per on the HR diagram
make it a highly diagnostic binary system to check
both helium content and overshooting treatment as well as to compare families
of stellar models in this transition region.
To this aim it would be worth to obtain an even more 
accurate atmospheric analysis (namely, individual abundances of the elements), from very high 
resolution (R$\sim$150\,000) spectroscopy. We plan to perform such an
investigation in the future. 

\begin{acknowledgements}
We would like to thank R.Barbon for assistance during the whole project,
and S.Cassisi for useful comments.
\end{acknowledgements}

\end{document}